\RequirePackage{snapshot} 
\documentclass[10pt,a4paper]{article}
\usepackage[utf8] {inputenc}
\usepackage{amsmath,amsfonts,amssymb}
\usepackage[dvips]{graphicx}
\usepackage{psfrag}
\usepackage{multicol}
\usepackage{algorithm}
\usepackage{fullpage}
\usepackage{authblk}
\usepackage{hyperref}

\usepackage{color}

\title{ \textbf{Simple Dynamics for Plurality Consensus}\thanks{Partially supported by  
   Italian MIUR-PRIN 2010-11  Project \emph{ARS TechnoMedia} and the EU FET Project \emph{MULTIPLEX} 317532. A preliminary version of this paper appeared in the \emph{Proceedings of the 26th ACM Symposium on Parallelism in Algorithms and Architectures (SPAA'14)}}}

\author[1]{L. Becchetti}
\author[2]{A. Clementi}
\author[1]{E. Natale}
\author[2]{F. Pasquale}
\author[1]{R. Silvestri}
\author[3]{L. Trevisan}

\affil[1]{\emph{Sapienza} Universit\`a di Roma, {\tt becchett@dis.uniroma1.it}, {\tt natale@di.uniroma1.it}, {\tt silvestri@di.uniroma1.it}   }
 
\affil[2]{Universit\`a \emph{Tor Vergata} di Roma, {\tt clementi@mat.uniroma2.it}, {\tt pasquale@mat.uniroma2.it},}

\affil[3]{Stanford University, {\tt trevisan@stanford.edu}}

\newtheorem{definition}{Definition}[section]

\newtheorem{lemma}[definition]{Lemma}

\newtheorem{theorem}[definition]{Theorem}
\newtheorem{cor}[definition]{Corollary}
\newtheorem{fact}{Fact}

\newcommand{\Prob}[2]{\mathbf{P}_{#1} \left( #2 \right)}
\newcommand{\probb}[1]{\mathbf{P}\left( #1 \right)}
\newcommand{\probc}[2]{\mathbf{P}\left(\left. #1 \;\right| #2 \right)}

\newcommand{\expec}[2]{\mathbb{E}\left[\left. #1 \;\right| #2 \right]}
\newcommand{\Expec}[2]{\mathbf{E}_{#1} \left[ #2 \right]}\newcommand{\proof}{\noindent\textit{Proof. }}
\newcommand{\pand}{\ensuremath{\wedge}}
\newcommand{\qed}{\hspace{\stretch{1}$\square$}}

\newcommand{\sM}{ {\mathcal M}^3}

\newcommand{\poly}{ {\mathrm{poly}}}
\newcommand{\polylog}{ {\mathrm{polylog} } }

\newcommand{\ideaproof}{\noindent\textit{Idea of the proof. }}

\renewcommand{\leq}{\leqslant}
\renewcommand{\le}{\leqslant}
\renewcommand{\geq}{\geqslant}
\renewcommand{\ge}{\geqslant}

\begin{document}
\thispagestyle{empty}
\maketitle

\begin{abstract}
We study a \emph{Plurality-Consensus} process in which each of $n$ anonymous 
agents of a communication network initially supports an opinion (a color chosen 
from a finite set $[k]$). Then, in every (synchronous) round, each agent can revise his color 
according to the opinions currently held by a random sample of his neighbors.
It is assumed that the initial color configuration exhibits a sufficiently large 
\emph{bias} $s$ towards a fixed plurality color, that is, the number of nodes 
supporting the plurality color exceeds the number of nodes supporting any other 
color by $s$ additional nodes.
The goal is having the process to converge to the \emph{stable} 
configuration in which all nodes support the initial plurality.
We consider a basic model in which the network is a clique and the update rule 
(called here the \emph{3-majority dynamics}) of the process is the following: each agent 
looks at the colors of three random neighbors and then applies the majority rule
(breaking ties uniformly). 

\smallskip
We prove that the process converges in time
$\mathcal{O}( \min\{ k, (n/\log n)^{1/3} \} \, \log n )$ 
with high probability, provided that 
$s \geqslant c \sqrt{ \min\{ 2k, (n/\log n)^{1/3} \}\, n \log n}$.
 We then prove that our upper bound above is tight as long as 
$k \leqslant (n/\log n)^{1/4}$. 
This fact implies an exponential time-gap between the plurality-consensus process
and the \emph{median} process studied by Doerr et al. in  [ACM SPAA'11].

\smallskip
A natural question is whether looking at more (than three) random neighbors can 
significantly speed up the process. We provide a negative answer to this 
question: In particular, we show that samples of polylogarithmic size can speed 
up the process by a polylogarithmic factor only. 


\end{abstract}
\newpage

\section{Introduction} \label{sec:intro}

We consider a communication network in which each of $n$ anonymous nodes
supports an initial opinion (a color chosen from a finite set $[k]$). In the
\emph{Plurality Consensus} problem, it is assumed that the initial (color)
configuration has a sufficiently large \emph{bias} $s$ towards a fixed color
$m \in [k]$ - that is, the number $c_m $ of nodes supporting the plurality color
(in short, the \emph{initial plurality size}) exceeds the number $c_j$ of nodes
supporting any other color $j$ by an additive value $s$ - and the goal is to
design an efficient fully-distributed protocol that lets the network converge to
the \emph{plurality consensus}, i.e., to the monochromatic configuration in
which all nodes support the plurality color. 

Reaching plurality consensus in a distributed system is a fundamental problem
arising in several areas such as Distributed Computing~\cite{DGMSS11,Pe02},
Communication Networks~\cite{PVV09}, and Social 
Networks~\cite{CDGNS13,MS10,MNT12}.
Inspired by some recent works analyzing simple updating-rules (called
\emph{dynamics}) for this problem~\cite{AD12,DGMSS11}, we study a discrete-time,
synchronous process in which, in every round, each of the $n$ anonymous
nodes revises his color according to a (small) random sample of neighbors.
We consider one of the simplest models, in which the network is a clique, and
the updating rule, called here \emph{$3$-majority dynamics}, is the 
following simple one: Each node
samples at random three neighbors, and picks the majority color among them
(breaking ties uniformly at random). We remark that looking at less than
three random neighbors would yield a coloring process that may converge to a
\emph{minority} color with constant probability even for $k=2$ and large initial
bias (i.e. $s = \Theta(n)$). 
   
In~\cite{DGMSS11}, a tight analysis of a $3$-neighbor dynamics for the
\emph{median} problem on the clique was presented: the goal there is to
converge to a stable configuration where all nodes support a  value which is a
good approximation of the \emph{median} of the initial configuration.
It turns out that, in the binary case (i.e $k=2$), the median problem is
equivalent to plurality consensus and the $3$-input dynamics for the
median is equivalent to the $3$-majority dynamics: As a result, they obtain, for
any bias $s \geqslant c \sqrt{n\log n}$ for some constant $c>0$, an optimal 
bound $\Theta(\log n)$ on the convergence time of the $3$-majority dynamics for
the binary case of the problem considered in this paper.

However, for any  $k \geqslant 3$, it is easy to see that the two problems above
differ significantly (in particular, the median may be very different from the
plurality) and thus, the two dynamics are different from each other as well.
Moreover, the analysis in~\cite{DGMSS11} - strongly based on the properties of
the median function - cannot be adapted to bound the convergence time of the
$3$-majority dynamics.
The role of parameter $k=k(n)$ in the convergence time of this dynamics is
currently unknown and, more generally, the existence of efficient dynamics
reaching plurality consensus for $k \geqslant 3$ is left as an important open
issue in~\cite{AAE07,DGMSS11,BD13}.  

\medskip
\noindent
\textbf{Our contribution.}
We present a new analysis of the $3$-majority dynamics in the general case (i.e.
for any $k \in [n] $). Our analysis shows that, with high probability (in short,
\emph{w.h.p.}\footnote{We say that a family of events $\{\mathcal{E}_n\}_n$
holds w.h.p. if a positive constant $c$ exists such that $\probb{\mathcal{E}_n}
\geqslant 1 - n^{-c}$ for sufficiently large $n$}), the process converges to
plurality consensus  within time
$\mathcal{O}\left( \min\{k, (n/\log n)^{1/3}\} \, \log n \right)$,
provided that the initial bias is
$s \geqslant c \sqrt{ \min\{2k, (n/\log n)^{1/3}\} \, n \log n}$, for some
constant $c>0$. 

Our proof technique is accurate enough to get another interesting form of the
above upper bound that does not depend on $k$. Indeed, when the initial
plurality size $c_m$ is larger than $n/ \lambda(n)$ for any function
$\lambda(n)$ such that $3  \leqslant \lambda(n) < \sqrt n$ and
$s \geqslant \sqrt{\lambda(n) \, n \log n}$,
then the process converges in time 
$\mathcal{O}\left(\lambda(n)\, \log n \right)$ w.h.p., no matter how large $k$
is.
Hence, when $c_m \geqslant n/\polylog (n)$ and
$s \geqslant \sqrt{n \, \polylog ( n)}$, the convergence time is polylogaritmic.

We then show that our upper bound is tight for a wide range of the input
parameters.
When $k \leqslant (n/\log n)^{1/4}$, we prove a lower bound
$\Omega(k \log n)$ on the convergence time of the $3$-majority dynamics starting
from some configurations with bias $s \leqslant (n/k)^{1-\epsilon}$, for an
arbitrarily small constant $\epsilon >0$.
Observe that this range largely includes the initial bias required by our upper
bound when $k \leqslant (n/\log n)^{1/4}$.
So, the \emph{linear-in-}$k$ dependence of the convergence time cannot be
removed for a wide range of the parameter $k$.

Our analysis also provides a clear picture of the $3$-majority dynamic process.
Informally speaking, the larger the initial value of $c_m$ is (w.r.t. $n$), the
smaller the required initial bias $s$ and the faster the convergence time are.
On the other hand, our lower-bound argument shows, as a by-product, that the
initial plurality size $c_m$ needs $\Omega(k\log n)$ rounds just to increase
from $n/k + o(n/k)$ to $2 \, n/k$.
Another natural issue is to analyze  the process under weaker assumptions on the initial bias.
We show that  there are initial configurations 
with  bias  $s = O(\sqrt{kn})$ for which the bias  
decreases in a single round with constant probability. This implies 
that under initial imbalances of this magnitude, it seems unlikely 
that one can prove upper bounds similar to ours above, at 
least with high probability.

We then prove a general negative result: Under the distributed model 
we consider, no dynamics with at most $3$ inputs (other than $3$-majority)
converges w.h.p. to plurality consensus starting from any initial 
configuration with bias $s=o(n)$.
In other words, not only can we not design a $3$-input 
dynamics that achieves convergence to plurality consensus in $o(k\log 
n)$ rounds, but the $3$-majority dynamics is the only one that 
eventually achieves this goal at all, no matter how long the process 
takes. 
Rather interestingly, by comparing the $\mathcal{O}(\log n)$ bound for the
median~\cite{DGMSS11} to  our negative results for the plurality
on the same distributed model, we get an exponential time-gap  between the 
task of computing the median and the one of computing plurality (this happens
for instance when $k = n^a$, for any constant $0< a< 1/4$).

A natural question suggested by our findings is whether (slightly) larger
random samples of nodes' neighborhoods might lead to significant 
improvements in convergence  time to plurality consensus. We provide a negative answer to this
question.
To this purpose, we consider $h$-plurality, i.e., the natural generalization of 
the $3$-majority dynamics in which every node, in each round, updates his color according to the
plurality of the colors supported by $h$ randomly sampled neighbors.
We prove a lower bound $\Omega \left( k/h^2 \right)$ on the convergence time of
the $h$-plurality dynamics, for integers $k$ and $h$ such that
$k/h = \mathcal{O}\left( n^{1/4 - \epsilon} \right)$, with $\epsilon$ an
arbitrarily-small positive constant.
We emphasize that scalable and efficient protocols must yield low communication
complexity and small node congestion in every round. These properties are
guaranteed by the $h$-plurality dynamics only when $h$ is small, say
$h = \mathcal{O}(\polylog (n))$: In this case, our lower  bound implies that the
resulting speed-up is only polylogarithmic with respect to the $3$-majority
dynamics.

One motivation for adopting dynamics in reaching (\emph{simple})
consensus\footnote{In the (simple) consensus problem the goal is to reach any
stable monochromatic configuration (any color is accepted) starting from any
initial configuration.}
(such as the median dynamics shown in~\cite{DGMSS11}) lies in their
provably-good \emph{self-stabilizing} properties against \emph{dynamic adversary
corruptions}: It turns out that the $3$-majority dynamics has good
self-stabilizing properties for the \emph{plurality consensus} problem.
More formally, a $T$-bounded adversary knows the state of every node at the end
of each round and, based on this knowledge, he can corrupt the color of up to
$T$ nodes in an arbitrary way, just before the next round begins.
In this case, the goal is to achieve an almost-stable phase where all but at
most $\mathcal{O}(T)$ nodes agree on the plurality value.
This ``almost-stability'' phase must have $\poly(n)$ length, with high
probability.
Our analysis implicitly shows that the $3$-majority dynamics guarantees the
self-stabilization property for plurality consensus for any $k$ and for
$T = o(s / k)$ if the initial bias is
$s \geqslant c \sqrt{ \min\{ 2k  , (n/\log n)^{1/3}  \} \, n \log n}$, for some
constant $c>0$.

\medskip
\noindent
\textbf{Related work.} The plurality consensus problem arises in several
applications such as distributed database management, where data redundancy or
replication and majority rules are used to manage the presence of unknown faulty
processors~\cite{DGMSS11,Pe02}.
The goal here is to converge to the version of the data supported by the 
plurality of the initial distributed copies (it is reasonable that a
sufficiently strong plurality of the nodes are not faulty and thus 
possess the correct data). Another application is distributed item
ranking, in which every node initially selects some item and the goal is to agree
on the most popular item according to the initial plurality opinion~\cite{PVV09}.
Further applications of majority updating rules in networks can be found
in~\cite{EK10,Pe02}.

The results most related  to our contribution are those in~\cite{DGMSS11} which
have been already discussed above.
Several variants of binary majority consensus have been studied in different
distributed models~\cite{AAE07,MS10}.

As for the \emph{population model}, where there is only one random node-pair
interaction per round (so the dynamics are strictly sequential), the binary case
on the clique has been analyzed in~\cite{AAE07} and their generalization to
the multivalued case ($k\geqslant 3$) does not converge to plurality even starting
with a large bias $s = \Theta(n)$.

The polling rule (a somewhat sequential-interaction version of the $1$-majority
dynamics) has been extensively studied on several classes of graphs
(see~\cite{Pe02}).

More expensive and complex protocols have been considered in order to speed up
the process. For instance, in~\cite{KT08}, a protocol for the
sequential-interaction model is presented that requires $\Theta(\log n)$ memory
per node and converges in time $\mathcal{O}(n^7)$. Other  protocols for the
sequential-interaction model have been analyzed in~\cite{BTV09,LB95} (with no
time bound).

In~\cite{BD13,DV12,PVV09}, the \emph{undecided-state} protocol  
on the continuous-time population model is proved to converge in
$\mathcal{O}(n \log n)$ expected time only for $k= \Theta(1)$ and
$s = \Theta(n)$: Even assuming such strong restrictions, the bound does not hold
in ``high probability'' and, moreover, their analysis, based on real-valued
differential-equations, do not work for the discrete-time parallel model
considered in this paper.  The simple rule  of the \emph{undecided-state} \cite{AAE07,PVV09} 
is to ``add'' one extra   state to somewhat account for the ``previous'' opinion supported by an agent.

In a recent work \cite{BCNPS15} (appeared after the conference version of this paper),
the undecided-state protocol has been analyzed on the discrete-time parallel model
for any $k = \mathcal{O}((n/\log n)^{1/3})$ and for initial 
configurations $\mathbf{c} = (c_1,\ldots, c_k)$
such that the  
(multiplicative) bias is  $c_m/c_j  = \Omega(1)$. There, it is 
shown that  this dynamics has  a convergence time which is w.h.p.  
linear in the \emph{monochromatic} distance of the initial 
configuration $\mathbf{c}$. The monochromatic distance of a 
configuration $\mathbf{c} = (c_1,\ldots, c_k)$ is defined as 
  
\[ 
	 \sum_{i=1}^k \, \left(\frac{c_i}{c_{m}}\right)^2.
\]
It turns out that there are initial configurations (in particular, those  having ``almost all''    nodes supporting 
only a polylogarithmic number of colors) from  which the undecided-state protocol is exponentially faster
than the $3$-majority. On the other hand, in addition to the  above condition on the multiplicative bias, we note
that, the undecided-state protocol may fail to reach consensus when $k = \omega(\sqrt n)$. 

Finally, protocols for specific network topologies and some ``social-based'' communities
have been studied in~\cite{AD12,DV12,MNT12,PVV09}. 

\medskip
\noindent
\textbf{Roadmap.} Section~\ref{sec::prely} formalizes the basic
concepts and gives some preliminary results.
Section~\ref{sec::UB} is devoted to the proofs of the upper bounds on the
convergence time of the $3$-majority dynamics.
In Section~\ref{sec::LB}, the lower bounds for the studied dynamics are
described.
Section~\ref{sec::conc} discusses some interesting open questions such as the
tightness of the initial bias. 

\section{Preliminaries} \label{sec::prely}
A \emph{($k$-color) configuration} (for short \emph{$k$-cd}) is any $k$-tuple $\mathbf{c} = (c_1,\ldots, c_k)$ such that $c_j$s are non negative integers and $\sum_{j=1,\ldots,k} c_j = n$.
In what follows, we will always assume wlog $c_1 \geqslant c_2 \geqslant \dots \geqslant c_k$. So $c_1$ is the \emph{plurality color} and $s(\mathbf c) = c_1 - c_2$ is the \emph{bias} of $\mathbf c$.

\smallskip\noindent
The $3$-majority protocol works as follows:
\begin{quote}
\textit{At every round, every node picks three nodes uniformly at random (including itself and with repetitions) and recolors itself according to the majority of the colors it sees.
If it sees three different colors, it chooses the first one.}
\end{quote}
\noindent
Clearly, in the case of three different colors, choosing the second or the third one would not make any difference.
The same holds even if the choice would be uniformly at random among the three colors.

\smallskip
For any round $t$ and for any $j \in [k]$, let $C^{(t)}_{j}$ be the r.v. counting the number of nodes colored $j$ at round $t$ and let $\mathbf{C}^{(t)} = (C^{(t)}_{1},\ldots,C^{(t)}_{k})$ denote the random variable indicating the $k$-cd at time $t$ of the execution of the $3$-majority protocol.

For every $j \in [k]$ let $\mu_j(\mathbf{c})$ be the expected number of nodes with color $j$ at the next round when the current $k$-cd is $\mathbf{c}$, i.e. $\mu_j(\mathbf{c}) = \expec{C^{(t+1)}_{j}}{\mathbf{C}^{(t)} = \mathbf{c}}$.

\begin{lemma}[Next expected coloring]\label{lemma:3maj-expected}
For any $k$-cd $\mathbf{c}$ and for every color $j\in [k]$, it holds that

\[
\mu_j(\mathbf{c})  = c_j\left[ 1 + \frac{1}{n^2}\left(nc_j - \sum_{h\in [k]} c_h^2\right)\right].
\]

\end{lemma}
\proof According to the $3$-majority protocol, a node $i$ gets color $j$ if it chooses three times color $j$, or if it chooses two times $j$ and one time a different color, or if it chooses the first time color $j$ and then, the second and third time, two different distinct colors. Hence if we name name $X_{i,j}^{(t)}$ the indicator random veriable of the event ``Node $i$ gets color $j$ at time $t$'', we have that
\begin{align*}
    P\left(X_{i,j}^{(t+1)}=1 \,|\, \mathbf{C}^{(t)} = \mathbf{c}\right)
    & =  \left(\frac{c_{j}}{n}\right)^{3}+3\left(\frac{c_{j}}{n}\right)^{2}\left(\frac{n-c_{j}}{n}\right)
    +\left(\frac{c_{j}}{n}\right)\left[1-\left(\frac{\sum_{h=1}^{k}c_{h}^{2}}{n^{2}}
    +2\left(\frac{c_{j}}{n}\right)\left(\frac{n-c_{j}}{n}\right)\right)\right]\\ 
    & =  \left(\frac{c_{j}}{n^{3}}\right)\left(n^{2}+c_{j}n-\sum_{h=1}^{k}c_{h}^{2}\right).
\end{align*}
\qed

\begin{lemma}[Next expected bias]
\label{lemma:3maj-amplification}
For any $k$-cd $\mathbf{c}$ and for every color $j \in [k]$ with $j \neq 1$, it holds that

\begin{equation}\label{eqlemma:3maj-amplification}
\mu_{1} - \mu_{j} \geqslant s(\mathbf c) \left(1+\frac{c_1}{n} \left(1- \frac{c_1}{n} \right)\right).
\end{equation}

\end{lemma}
\begin{proof}
Observe that, when we assume $c_1 \geqslant c_2 \geqslant \dots \geqslant c_k$, we can give the following upper bound on the sum of squares in Lemma~\ref{lemma:3maj-expected}
\begin{equation}\label{eq:ubsumsquares}
\sum_{h \in [k]} c_h^2 
= c_1^2 + \sum_{h = 2}^k c_h^2
\leqslant c_1^2 + c_2 \sum_{h = 2}^k c_h
\leqslant c_1^2 + n c_2.
\end{equation}
From Lemma~\ref{lemma:3maj-expected} it thus follows that, for any $j \neq 1$, 
\begin{align*}
\mu_1 - \mu_j \geqslant \mu_1 - \mu_2 
&= (c_1 - c_2) + \frac{\left( c_1^2 - c_2^2 \right)}{n} - \frac{c_1 - c_2}{n^2} \sum_{h \in k} c_h^2 \\
& =  s(\mathbf c) \left( 1 + \frac{c_1 + c_2}{n} - \frac{1}{n^2} \sum_{h \in k} c_h^2 \right) \\
& \geqslant s(\mathbf c) \left( 1 + \frac{c_1 + c_2}{n} - \frac{c_1^2 + n c_2}{n^2} \right) \\
& =  s(\mathbf c) \left( 1 + \frac{c_1}{n}\left( 1 - \frac{c_1}{n} \right) \right),
\end{align*}
where in the inequality we used~\eqref{eq:ubsumsquares} and the fact that $c_1 - c_2 \geqslant 0$.
\qed
\end{proof}

\section{Upper bounds for 3-majority}\label{sec::UB}
In this section, we provide  the following  upper bound on the convergence time of the $3$-majority dynamics  
which clarifies the roles played by  the plurality color and by  the initial bias.

\begin{theorem}[the general upper bound] \label{theo:3-majority-ub}
    Let $\lambda$ be any value such that $ \lambda <  \sqrt[3]{n}$ and
let $\mathbf{c}$ be any initial $k$-cd, with $c_1 \geqslant n / \lambda$ and
$s(\mathbf{c}) \geqslant 24\sqrt{ 2 \lambda \, n \log n}$. Then the
$3$-majority protocol converges to the plurality color in
$\mathcal{O}\left(\lambda \, \log n \right)$ time w.h.p.
\end{theorem}

\smallskip\noindent
The next three corollaries of Theorem~\ref{theo:3-majority-ub}  address three relevant special cases.
Corollary~\ref{cor:3-majority-ub-k} is obtained by setting $\lambda = \min\left\{ 2k,\; \sqrt[3]{{n} / {\log n}}\right\}$ and it provides a bound which does not assume any condition on $c_m$.
 
 \begin{cor}\label{cor:3-majority-ub-k}
Let $\mathbf{c}$ be any initial $k$-cd with
\[
s(\mathbf{c}) \geqslant 22\sqrt{\min\left\{2k,\; \sqrt[3]{\frac{n}{\log n}}\right\}n\log n}.
\]
Then, the $3$-majority protocol converges to the plurality color in $\mathcal{O}\left(\min\left\{2k,\; \sqrt[3]{{n} / {\log n}}\right\}\log n\right)$ time w.h.p.
\end{cor}

 \smallskip\noindent
Corollaries~\ref{cor:3-majority-ub-log}  and~\ref{cor:3-majority-ub-const}
are  obtained by setting  $\lambda =  \poly\log(n)$ and $\lambda = \Theta(1)$, respectively.    They  
 provide sufficient conditions for a polylogarithmic convergence time.

\begin{cor}\label{cor:3-majority-ub-log}
Let $\mathbf{c}$ be any initial $k$-cd with $c_1 \geqslant n/\log^\ell n$ and $s(\mathbf{c}) \geqslant 22\sqrt{n\log^{\ell + 1} n}$. Then, the $3$-majority protocol converges to the plurality color in $\mathcal{O}(\log^{\ell + 1} n)$ time w.h.p.
\end{cor}

\begin{cor}\label{cor:3-majority-ub-const}
Let $\mathbf{c}$ be any $k$-cd with $c_1 \geqslant  n/ \beta$ and 
$s(\mathbf{c}) \geqslant 22\sqrt{\beta n\log n}$, for some constant $\beta \geqslant 3$. Then,
the $3$-majority protocol converges to the plurality color in $\mathcal{O}(\log n)$ time w.h.p.
\end{cor}

   In order to prove Theorem~\ref{theo:3-majority-ub},  we need the following three  technical lemmas that essentially characterize three different phases of
  the process analysis. Each of them concerns a different range assumed by the plurality  $c_1$.
  The first lemma considers configurations in which $c_1$ is under a suitable 
  constant fraction of $n$: in this case, it shows  that the bias between the plurality size and the 
  size of any other color increases by a factor $1+\Omega(c_1/n)  = 1 + \Omega(1/\lambda)$. 

\begin{lemma}[from plurality to majority]\label{lemma:plurtomaj}
Let $\mathbf{c}$ be any $k$-cd with ${n} / {\lambda}
\leqslant c_1 \leqslant {2n} / {3}$ and $s(\mathbf c) \geqslant
\alpha\sqrt{\lambda n\log n}$ where $ \lambda <  \sqrt[3]{n}$ 
and $\alpha$ is a sufficiently large constant. 
Then, 
for any other color $j \neq 1$ it holds that
\begin{equation}\label{eq:ampl_bias}
\Prob{}{ \left. C_1^{(t+1)} - C_j^{(t+1)} \geqslant
    s(\mathbf{c})\left(1+\frac{c_{1}}{4n}\right) \,\right| \, \mathbf{C}^{(t)} =
    \mathbf{c}} 
\geqslant 1 - \frac 1{n^{3}}.
\end{equation}
\end{lemma}
\begin{proof}
    \newcommand{\constone}{3}
Conditional on any configuration $\mathbf{c}$, from the Chernoff bound w.h.p. 
it holds that
\begin{align}
    \label{eq:upper_c2}
    C_{j} &\leqslant \max \left\{ \mu_{j}+\alpha\sqrt{\mu_{j}\log n}, \log n \right\}\\
    C_{1} &\geqslant \mu_{1} - \alpha\sqrt{\mu_{1}\log n} \nonumber 
\end{align}
Thus, if 
$\mu_{j}+\alpha\sqrt{\mu_{j}\log n} \geqslant \log n$,
w.h.p. it holds that\footnote{We are using the fact that 
    $\Pr\left(A\cap B\right)\geq1-\Pr\left(A^{C}\right)-\Pr\left(B^{C}\right)$.} 
\begin{equation}\label{eq:whp_bias}
C_{1}-C_{j} \geqslant \mu_{1}-\mu_{j}-\alpha\sqrt{\mu_{1}\log n}-\alpha\sqrt{\mu_{j}\log n}
\geqslant \mu_{1}-\mu_{j}-2\alpha\sqrt{\mu_{1}\log n}.
\end{equation}
Otherwise, if 
$\mu_{j}+\alpha\sqrt{\mu_{j}\log n} < \log n$, then w.h.p. it holds that
\begin{align}
    \label{eq:whp_bias_smallc2}
    C_{1}-C_{j} &\geq \mu_{1}-\alpha\sqrt{\mu_{1}\log n}-2\log n \nonumber \\
    &\geq \mu_{1}- \frac 43 \alpha\sqrt{\mu_{1}\log n} \nonumber \\
    &\geq \mu_{1}- \frac 53 \alpha\sqrt{\mu_{1}\log n} 
        - \mu_{j}- \alpha \sqrt{\mu_{j}\log n} \nonumber \\
    &\geqslant \mu_{1}-\mu_{j}-2\alpha\sqrt{\mu_{1}\log n}.
\end{align}

From Lemma~\ref{lemma:3maj-amplification} and the hypothesis $c_{1} \leqslant {2n} / {3}$
we get that
\[
\mu_{1}-\mu_{j} \geqslant \left(c_{1}-c_{j}\right)\left(1+\frac{c_{1}}{3n}\right).
\]
Thus, in (\ref{eq:whp_bias}) and (\ref{eq:whp_bias_smallc2}) we get
\begin{align*}
\mu_{1} - \mu_{j} - 2\alpha\sqrt{\mu_{1}\log n} 
& \geqslant \left(c_{1}-c_{j}\right)\left(1+\frac{c_{1}}{3n}\right)-2\alpha\sqrt{2c_{1} \log n}\\
& \geqslant \left(c_{1}-c_{j}\right)\left(1+\frac{c_{1}}{3 n}\right) - 2\alpha\sqrt{2c_{1} \log n} \\
& \geqslant \left(c_{1}-c_{j}\right)
    \left(1+\frac{c_{1}}{3n}-\frac{2\alpha\sqrt{2c_{1} \log n}}{\left(c_{1}-c_{j}\right)}\right) \\
& \geqslant \left(c_{1}-c_{j}\right)\left(1+\frac{c_{1}}{3n}-\frac{\sqrt{c_{1}/n}}{12\sqrt{\lambda}}\right) \\
& \geq  \left(c_{1}-c_{j}\right)
    \left(1+\frac{c_{1}}{3 n}\left(1-\frac{1}{4\sqrt{c_{1}\lambda /n}}\right)\right) \\
& \geqslant \left(c_{1}-c_{j}\right)\left(1+\frac{c_{1}}{4 n}\right),
\end{align*}
concluding the proof. 

\smallskip\noindent
 
\qed
\end{proof}

\noindent
Once $c_1$ becomes larger than $2n/3$ the negative occurrence of $c_1$ in  (\ref{eqlemma:3maj-amplification}) does not allow to 
directly show a drift towards plurality.  We thus  consider another  useful  ``drift''  of the process:  
The sum  of  all the other color sizes decreases exponentially  w.h.p.,  as long as this sum  is enough large to apply concentration bounds.
This result is formalized in the next lemma.

\begin{lemma}[from majority to almost all]\label{lemma:majtoaall}
Let $\mathbf{c}$ be any $k$-cd with ${2n} / {3}\leq c_{1} \leqslant n-\omega\left(\log n\right)$.
Then, it holds that
\[
\Prob{}{\left. \sum_{i\neq1} C_{i}^{(t+1)} \leqslant \frac{8}{9} \sum_{i\neq1} c_{i} \, \right| \, \mathbf{C}^{(t)} = \mathbf{c} } 
\geqslant 1 - \frac 1 {n^{3}}.
\]
\end{lemma}
\begin{proof}
Let us define $\mu_{-1}^{\left(t\right)}=\sum_{i\neq1}\mu_{i}^{\left(t\right)}$.
By using (\ref{eq:ubsumsquares}) we have 
\begin{align*}
    \frac{\mu_{-1}^{\left(t+1\right)}}{n}
    &= \sum_{i\neq1}\frac{c_{i}}{n}\left(1+\frac{c_{i}}{n}
        - \sum_j \left( \frac{ c_{j} }n \right)^2\right)\\
    &= 1-\frac{c_{1}}{n}+\sum_{i\neq1}\left( \frac{c_{i}}{n} \right)^{2}
        -\left(1-\frac{c_{1}}{n}\right) \sum_j \left( \frac{ c_{j} }n \right)^2 \\
    &= 1-\frac{c_{1}}{n}-\left( \frac{c_{1}}{n} \right)^{2}
        +\frac{c_{1}}{n} \sum_j \left( \frac{ c_{j} }n \right)^2\\
    &\leq 1-\frac{c_{1}}{n}-\left( \frac{c_{1}}{n} \right)^{2}
        +\frac{c_{1}}{n}\left( \left( \frac{ c_{1} }{n} \right)^{2}
        + \frac{ c_{2} }{n}\left(1- \frac{ c_{1} }{n}\right)\right)\\
    &= \left(1- \frac{ c_{1} }{n}\right)
        \left(1- \left( \frac{ c_{1} }{n} \right)^{2}
        + \frac{ c_{1} }{n} \frac{ c_{2} }{n}\right)
    = \left(1- \frac{ c_{1} }{n}\right)
        \left(1- \frac{ c_{1} }{n}\left( \frac{ c_{1} }{n}
        - \frac{ c_{2} }{n}\right)\right).
\end{align*}
Using the hypothesis $ { c_{1} } / {n} \geq {2} / {3}$ 
(hence $c_{2} / n \leq {1} / {3}$)
the last expression become
\begin{equation}
    \label{eq:expected_nonmaj}
    \left(1- \frac{ c_{1} }{n}\right)\left(1- \frac{ c_{1} }{n}\left( \frac{ c_{1} }{n}- \frac{ c_{2} }{n}\right)\right)
    \leq \left(1- \frac{ c_{1} }{n}\right)\left(1-\frac{ c_{1} }{3n}\right)
    \leq \left(1- \frac{ c_{1} }{n}\right)=\frac{7}{9}\frac{\mu_{-1}^{\left(t\right)}}{n}.
\end{equation}
Now observe that, from the Chernoff bound, 
as long as $\mu_{-1}^{(t+1)} \in \omega\left(\log n\right)$,
w.h.p. it holds 

\begin{equation}\label{eq:upper_on_sum}
\sum_{i\neq1}C_{i}^{\left(t+1\right)}\leq\mu_{-1}^{\left(t+1\right)}
    +\sqrt{\mu_{-1}^{\left(t+1\right)}\log n}
    =\mu_{-1}^{\left(t+1\right)}
    \left(1+ \sqrt{\frac{\log n}{\mu_{-1}^{\left(t+1\right)}}}\right)
    =\mu_{-1}^{\left(t+1\right)}\left(1+o\left(1\right)\right).
    \end{equation}
    Moreover, from Lemma~\ref{lemma:3maj-expected} it follows that 
\begin{equation}\label{eq:upper_on_expected}
\mu_{1} \leqslant 2 c_{1}.
\end{equation}
Thus, by replacing (\ref{eq:upper_on_expected}) and (\ref{eq:expected_nonmaj})
in (\ref{eq:upper_on_sum}), we get that w.h.p. it holds 
\[
\sum_{i\neq1}C_{i}^{\left(t+1\right)}\leq\frac{7}{9}\mu_{-1}^{\left(t+1\right)}\left(1+o\left(1\right)\right)\leq\frac{8}{9}\sum_{i\neq1}\mu_{i}^{\left(t+1\right)},
\]
concluding the proof.

\smallskip\noindent
 
\qed
\end{proof}

\noindent
Finally, when  the sum of all  the  minority colors  is not larger than a polylogarithmic function,   the probability that
they all disappear in one round  is high. This is shown in the new, final lemma.

\begin{lemma}[the last step]\label{lemma:laststep}
Let $\alpha>0$ and let $\mathbf{c}$ be any $k$-cd with $c_{1} \geqslant n-\log^{\alpha}n$. Then, it holds that 
\begin{equation}\label{eq:nonmaj_disappear}
	\Prob{}{\sum_{i \neq 1}C_{i}^{(t+1)} = 0 \,|\, \mathbf{C}^{(t)} = \mathbf{c}} \geqslant 1 -  \frac{3\log^{2\alpha}n}{n}.
\end{equation}
\end{lemma}
\begin{proof}
As in the previous proof let 
$\mu_{-1} = \sum_{i\neq1}\mu_{i}$.
Note that 
$c_{1} \geq n-\log^{\alpha}n$ implies $\sum_{i\neq1}c_{i}\leq\log^{\alpha}n$.
Thus, from Lemma~\ref{lemma:3maj-expected} we have 
\begin{align*}
    \mu_{-1}
    &=\sum_{i\neq1}c_{i}\left(1+ \frac{ c_{i} }{n}
        -\sum_j \left( \frac{ c_{j} }n \right)^2\right)\\
    &\leq \sum_{i\neq1}c_{i}\left(1+ \frac{ c_{i} }{n}
        - \left( \frac{ c_{1} }{n} \right)^{2}\right)\\
    &= \sum_{i\neq1}c_{i}\left(1+ \frac{ c_{i} }{n}
        -\left(1-\sum_{j\neq1} \frac{ c_{j} }{n}\right)^{2}\right)\\
    &\leq \sum_{i\neq1}c_{i}\left( \frac{ c_{i} }{n}
        +2\sum_{j\neq1} \frac{ c_{j} }{n}\right)\\
    &\leq \sum_{i\neq1}c_{i}\left(\frac{3\log^{\alpha}n}{n}\right) 
    = \frac{3\log^{2\alpha}n}{n}.
\end{align*}
Finally, (\ref{eq:nonmaj_disappear}) follows from Markov's inequality
on the event ``$\sum_{i\neq1}C_{i}^{\left(t+1\right)}\geq1$'' and, since
$\sum_{i\neq1}C_{i}^{\left(t+1\right)}$ is a non-negative integer-valued r.v., 
this is equivalent as ``$\sum_{i\neq1}C_{i}^{\left(t+1\right)}>0$''.

\smallskip\noindent
 
\qed
\end{proof}

\paragraph{Proof of Theorem \ref{theo:3-majority-ub}}
From Lemma~\ref{lemma:plurtomaj}, we prove that, as long as the number of nodes with
the plurality color $c_1$ is smaller than a constant fraction of $n$, the bias
between $c_1$ and $c_2$ increases by a factor $(1 + \lambda^{-1})$, w.h.p. (ii)
In Lemma~\ref{lemma:majtoaall} we prove that, when the plurality color reaches a
suitable constant fraction of $n$, then the number of nodes with non-plurality
colors decreases at exponential rate, w.h.p. Finally, (iii) in
Lemma~\ref{lemma:laststep} we consider separately the last step of the
protocol, where all colors but the plurality one disappear w.h.p. \qed

\paragraph{Plurality consensus in an  adversarial model.}
Let us consider a dynamic adversary that can change the color of up to $T$
nodes at the beginning of each round  and assume
 $T = o( s / \lambda)$.
Then, Theorem~\ref{theo:3-majority-ub} still holds since the impact of such a
\emph{$T$-bounded } adversary is negligible in the growth of the bias $s$. 
Indeed, from Lemma \ref{lemma:plurtomaj} w.h.p. it holds
\begin{equation*}
    C_1^{(t+1)} - C_j^{(t+1)} \geqslant s(\mathbf{c}) + \frac{s(\mathbf{c})}{4\lambda}
\end{equation*}
Then, for any $T = o( s / \lambda)$, w.h.p. we have that
\begin{equation*}
    C_1^{(t+1)} - C_j^{(t+1)} \geqslant s(\mathbf{c}) + \frac{s(\mathbf{c})}{4\lambda}
        - T \geq s(\mathbf{c}) + \Theta\!\left( \frac{s(\mathbf{c})}{\lambda} \right)
\end{equation*}

For instance, from Corollary \ref{cor:3-majority-ub-k}, 
when $k \leqslant 2 \sqrt[3]{{n} / {\log n}}$, the
resilience of the 3-majority dynamics is $T= o(s/k)$.

\section{Lower bounds}\label{sec::LB}
This section is organized in four subsections: in the first one, we 
prove a lower bound on the convergence time of the $3$-majority 
dynamics; in the second subsection, we show that $3$-majority is 
essentially the only $3$-input dynamics that converges to plurality 
consensus; in the third subsection, we provide a lower 
bound on the convergence time of the $h$-plurality dynamics for $h > 
3$; finally, in the fourth subsection we show that our assumption on 
the magnitude of the initial bias is in a sense (almost) tight if one 
wants to prove the bounds of Section \ref{sec::UB} with high 
probability.

\subsection{Lower bound for 3-majority}\label{sec::LBM}
In this section we show that if the $3$-majority dynamics starts from a sufficiently balanced configuration (i.e., at the beginning there are $n/k\pm o(n/k)$ nodes of every color) then it will take $\Omega(k \log n)$ rounds w.h.p. to reach one of the absorbing configurations where all nodes have the same color.
In what follows, all events and random variables thus concern the Markovian process yielded by the $3$-majority dynamics.

In the next lemma we show that if there are at most $n/k + b$ nodes of a specific color, where $b$ is smaller than $n/k$, then at the next round there are at most $n/k + (1 + 3/k) b$ nodes of that color w.h.p.

\begin{lemma}\label{lemma:lbincreasingrate}
Let the number of colors $k$ be such that $k \leqslant \left(n/\log n \right)^{1/4}$, let $b$ be any number with $k \sqrt{n \log n} \leqslant b \leqslant n/k$, and let $\mathbf{c} = (c_1, \dots, c_k)$ be a configuration. If $c_j = n/k + a$ for some color $j \in [k]$ and for some $a \leqslant b$, then the number of nodes with color $j$ at the next round are at most $n/k + (1+3/k)b$ w.h.p.; more precisely, for any $a \leqslant b$ and for any configuration $\mathbf{c}$ such that $c_j = n/k + a$ it holds that
\[
\Prob{}{C_j^{(t+1)} \geqslant \left.\frac{n}{k} + \left( 1 + \frac{3}{k} \right) b \;\right|\; \mathbf{C}^{(t)} = \mathbf{c}} \leqslant \frac{1}{n^2}.
\]  
\end{lemma}
\proof
For any configuration $\mathbf{c} = (c_1, \dots, c_k)$ with $\sum_{j=1}^k c_j = n$ and any color $j \in [k]$, the expected value of the number of nodes colored $j$ at round $t+1$ conditional on $\left\{\mathbf{C}^{(t)} = \mathbf{c} \right\}$ is (see Lemma~\ref{lemma:3maj-expected})
\[
\Expec{}{C_j^{(t+1)} \; | \; \mathbf{C}^{(t)} = \mathbf{c}} = c_{j} \left(1 + \frac{c_{j}}{n} - \frac{1}{n^2} \sum_{j=1}^k c_j^2 \right).
\]
Observe that, since $\sum_{j=1}^k c_j = n$, from Jensen inequality (see Lemma~\ref{lemma:jensen}) it follows that $(1/n^2)\sum_{j=1}^k c_j^2 \geqslant 1/k$. Hence, we can give an upper bound on the expectation of $C_j^{(t+1)}$ that depends only on $c_j$ and not on the whole configuration $\mathbf{c}$ at round $t$, namely
\[
	\Expec{}{C_j^{(t+1)} \,|\, \mathbf{C}^{(t)}} \leqslant C_j^{(t)} \left( 1 + \frac{C_j^{(t)}}{n} - \frac{1}{k} \right).
\]
If we condition on the number of nodes of color $j$ being $c_j = n/k + a$ in configuration $\mathbf{c}$, for some $a \leqslant b$, we get
\begin{multline*}
\Expec{}{C_j^{(t+1)} \,|\, \mathbf{C}^{(t)} = \mathbf{c}} \leqslant 
    \left( \frac{n}{k} + a \right) \left( 1 + \frac{n/k + a}{n} - \frac{1}{k}
    \right)
 =  \frac{n}{k} + \left( 1 + \frac{1}{k} \right) a + \frac{a^2}{n} \\
 \leqslant  \frac{n}{k} + \left( 1 + \frac{1}{k} \right) b + \frac{b^2}{n} 
\leqslant  \frac{n}{k} + \left( 1 + \frac{2}{k} \right) b,
\end{multline*}
where in the last two inequalities we used that $a \leqslant b$ and $b
\leqslant n/k$.\footnote{Notice that the inequality holds in particular for
negative $a$ as well} Since $C_j^{(t+1)}$ conditional on
$\left\{\mathbf{C}^{(t)} = \mathbf{c}\right\}$ can be written as a sum of $n$
independent Bernoulli random variables, from the Chernoff bound (see
Lemma~\ref{lemma:cb}) we thus get that for every $a \leqslant b$ it holds that 
\begin{equation*}
	\Prob{}{\left. C_j^{(t+1)} \geqslant \frac{n}{k} + \left( 1 + \frac{3}{k} \right) b \, \right| \, \mathbf{C}^{(t)} = \mathbf{c}}  \leqslant e^{-2 (b/k)^2 / n}  \leqslant \frac{1}{n^2},
\end{equation*}
where in the last inequality we used that $b \geqslant k \sqrt{n \log n}$.
\qed

\medskip\noindent
Let us say that a configuration $\mathbf{c} = (c_1, \dots, c_k) \in \{ 0, 1,
\dots, n \}^k$ with $\sum_{j=1}^k c_j = n$ is \emph{monochromatic} if there is
an $j \in [k]$ such that $c_j = n$. In the next theorem we show that if we
start from a sufficiently \emph{balanced} configuration, then the $3$-majority
protocol takes $\Omega(k \log n)$ rounds w.h.p. to reach a monochromatic
configuration. 

\begin{theorem}\label{theorem:lowerbound}
Let $\tau = \inf\{ t \in \mathbb{N} \,:\, \mathbf{C}^{(t)} \mbox{ is monochromatic} \}$
be the random variable indicating the first round such that the system is in a
monochromatic configuration. If the initial number of colors is $k \leqslant
(n/\log n)^{1/4}$ and the initial configuration is $\mathbf{c} = (c_1, \dots,
c_k)$ with $\max \{ c_j \,:\, j = 1, \dots, k \} \leqslant \frac{n}{k} + \left(
\frac{n}{k} \right)^{1-\varepsilon}$ then $\tau = \Omega(k \log n)$ w.h.p.  
\end{theorem}
\ideaproof 
For a color $j \in [k]$ let us denote the difference $C_{j} - n/k$  as  the
\emph{positive unbalance}. In Lemma~\ref{lemma:lbincreasingrate} we proved
that, as long as the positive unbalance of a color is smaller than $n/k$, this
will increase by a factor smaller than $(1 + 3/k)$   at every round (w.h.p.).
Hence, if a color starts with a positive unbalance smaller than
$(n/k)^{1-\varepsilon}$, then it will take $\Omega(k \log n)$ rounds to reach
an unbalance of $n/k$ w.h.p. By union bounding on all the colors, we can get
the stated  lower bound. 
\qed
\smallskip

\proof
Observe that if $T \leqslant c \, k \log n$, for a suitable positive constant
$c$, then $(1 - 3/k)^T (n/k)^{1-\varepsilon}$ is smaller than $n/k$. Since in
the initial configuration $\mathbf{c}$ for any color $j \in [k]$ we have that
$c_j \leqslant {n} / {k} + \left( {n} / {k} \right)^{1-\varepsilon}$, for $T \leqslant c \,
k \log n$ it holds that

\begin{equation}\label{eq:upton}
\Prob{}{C_j^{(t)} = n \,\left|\, \mathbf{C}^{(0)} = \mathbf{c} \right.}
\leqslant
\Prob{}{C_j^{(T)} \geqslant \frac{n}{k} + \left(1 + \frac{3}{k}\right)^T \left(
\frac{n}{k} \right)^{1-\varepsilon} \Big|\, \mathbf{C}^{(0)} = \mathbf{c} },
\end{equation}

\noindent Since 
$c_j \leqslant {n} / {k} + \left( {n} / {k} \right)^{1-\varepsilon}$, if we
also have 
$C_j^{(T)} \geqslant {n} / {k} + \left(1 + {3} / {k}\right)^T
\left( {n} / {k} \right)^{1-\varepsilon}$, then a round $t$ with 
$0 \leqslant t \leqslant T-1$ must exist such that 
$C_j^{(t)} \leqslant n/k + b$ and
$C_j^{(t+1)} \geqslant n/k + (1 + 3/k)b$ for some value $b$, with 
$k \sqrt{n \log n} \leqslant b \leqslant n/k$, thus

\begin{align}
&\Prob{}{C_j^{(T)} \geqslant \frac{n}{k} + \left(1 + \frac{3}{k}\right)^T
   \left( \frac{n}{k} \right)^{1-\varepsilon} \Big|\, \mathbf{C}^{(0)} =
   \mathbf{c} } 
   \label{eq:first_lower} \\
& \leqslant \Prob{}{\exists 0 \leqslant t \leqslant T-1 \;:\; C_j^{(t)}
   \leqslant \frac{n}{k} + b \, \mbox{ and } \, C_j^{(t+1)} \geqslant
   \frac{n}{k} + \left(1 + \frac{3}{k}\right) b \;\left|\; \mathbf{C}^{(0)} =
   \mathbf{c} \right.} 
   \label{eq:second_lower} \\
& \leqslant \sum_{t=0}^{T-1} \Prob{}{C_j^{(t)} \leqslant \frac{n}{k} +
    b_t \, \mbox{ and } \, C_j^{(t+1)} \geqslant \frac{n}{k} + \left(1 +
    \frac{3}{k}\right) b_t \;\left|\; \mathbf{C}^{(0)} = \mathbf{c} \right. }
    \label{eq:third_lower}
\end{align}
where the inequality from \eqref{eq:first_lower} to \eqref{eq:second_lower} 
holds for some $b$ with 
$k \sqrt{n \log n} \leqslant b \leqslant n/k$, and the inequality 
from \eqref{eq:second_lower} to \eqref{eq:third_lower} holds for some 
$b_0, \dots, b_{T-1}$ with 
$k \sqrt{n \log n} \leqslant b_t \leqslant n/k$ for every $t=0,\dots, T-1$.
Now observe that
\begin{align}\label{eq:technicalities}
&\Prob{}{C_j^{(t)} \leqslant \frac{n}{k} + b_t \, \mbox{ and } \, C_j^{(t+1)}
   \geqslant \frac{n}{k} + \left(1 + \frac{3}{k}\right) b_t \;\Big|\;
   \mathbf{C}^{(0)} = \mathbf{c}}\nonumber\\
& = \sum_{a \leqslant b_t} \Prob{}{C_j^{(t)} = \frac{n}{k} + a \, \mbox{ and }
   \, C_j^{(t+1)} \geqslant \frac{n}{k} + \left(1 + \frac{3}{k}\right) b_t
   \;\left|\; \mathbf{C}^{(0)} = \mathbf{c} \right. } \nonumber \\
& = \sum_{a \leqslant b_t} \Prob{}{ C_j^{(t+1)} \geqslant \frac{n}{k} +
   \left(1 + \frac{3}{k}\right) b_t \;\left|\; C_j^{(t)} = \frac{n}{k} + a \mbox{
   and } \mathbf{C}^{(0)} = \mathbf{c} \right. } 
   \cdot \Prob{}{C_j^{(t)} = \frac{n}{k} + a \;\Big|\; \mathbf{C}^{(0)} =
   \mathbf{c} } \nonumber \\
& \leqslant \frac{1}{n^2} \sum_{a \leqslant b_t} \Prob{}{C_j^{(t)} =
    \frac{n}{k} + a \;\Big|\; \mathbf{C}^{(0)} = \mathbf{c} } \leqslant
    \frac{1}{n^2},
\end{align}
where in the last line we used Lemma~\ref{lemma:lbincreasingrate}.

By combining (\ref{eq:upton}), (\ref{eq:third_lower}), and
(\ref{eq:technicalities}) we get that, for every color $j \in [k]$, if the
initial number of nodes colored $j$ is $c_j \leqslant {n} / {k} +
\left({n} / {k}\right)^{1-\varepsilon}$ at any round $T \leqslant c \, k \log
n$ the probability that all nodes are colored $j$ is at most $T/n^2$. The
probability that $\mathbf{C}^{(T)}$ is monochromatic is thus at most $(kT)/n^2
\leqslant n^{-\alpha}$ for some positive constant $\alpha$.
\qed

\medskip\noindent
It may be worth noticing that what we actually prove in Theorem~\ref{theorem:lowerbound} is that $\Omega(k \log n)$ rounds are required in order to go from a configuration where the majority color has at most $n/k + (n/k)^{1-\varepsilon}$ nodes to a configuration where it has $2n/k$ colors.

\subsection{A negative result for 3-input dynamics} \label{sec:LBgeneral}
In order to prove that dynamics that differ from the majority ones do not solve plurality consensus, we first give some formal definitions of the dynamics we are considering.  

\begin{definition}[$\mathcal{D}_h(k)$ protocols]
An \emph{$h$-dynamics} is a synchronous protocol where at each round every node picks $h$ random neighbors (including itself and with repetition) and recolors itself according to some deterministic rule that depends only on the colors it sees.
Let $\mathcal{D}_h(k)$ be the class of $h$-dynamics and observe that a dynamics $\mathcal{P} \in \mathcal{D}_h$ can be specified by a function
\[ 
f\,:\, [k]^h \rightarrow [k],
\]
such that $f(x_1, \dots, x_h) \in \{ x_1, \dots, x_h \}$. Where $f(x_1, \dots, x_h)$ is the color chosen by a node that sees the (ordered) sequence $(x_1, \dots, x_h)$ of colors.
\end{definition}

\noindent
In the class $\mathcal{D}_3(k)$, there is a subset $\sM$ of equivalent protocols called $3$-majority dynamics having two key-properties
described below: the clear-majority   and the uniform  one. 

\begin{definition}[clear-majority property] \label{def:maj}
Let $(x_1,x_2,x_3) \in [k]^3$ be a triple of colors. We say that $(x_1,x_2,x_3)$ has a \emph{clear majority} if at least two of the three entries have the same value.
A dynamics $\mathcal{P} \in \mathcal{D}_3(k)$ has the \emph{clear-majority} property if whenever its $f$ sees a clear majority it returns the majority color.
\end{definition}

\noindent
Given any $3$-input dynamics function 
$f(x_1,x_2,x_3)$, for any triple of distinct colors $r,g,b \in [k]$, let $\Pi(r,g,b)$ be the subset of permutations of the colors $r,g,b$ and define the following ``counters'':
\begin{align*}
\delta_r & =  |\{(z_1,z_2,z_3) \in \Pi(r,g,b) , \ s.t. \ f(z_1,z_2,z_3) = r \}|, \\
\delta_g & =  |\{(z_1,z_2,z_3) \in \Pi(r,g,b) , \ s.t. \ f(z_1,z_2,z_3) = g \}|, \\
\delta_b & =  |\{(z_1,z_2,z_3) \in \Pi(r,g,b) , \ s.t. \ f(z_1,z_2,z_3) = b \}|.
\end{align*}

\noindent
Observe that for any $3$-inputs dynamics it must hold $\delta_g + \delta_r + \delta_b = 6$.

\begin{definition}[uniform property]
A dynamics $\mathcal{P} \in \mathcal{D}_3(k)$ has the \emph{uniform} property if, for any triple of distinct colors $r,g,b \in [k]$, it holds that $\delta_r = \delta_g = \delta_b \ (= 2)$.
\end{definition}

\noindent
Informally speaking, the clear-majority and the uniform properties provide a clean characterization of those dynamics that are good solvers for plurality consensus.
This fact is formalized in the next definitions and in the final theorem.

\begin{definition} [$3$-majority dynamics]
A protocol $\mathcal{P} \in \mathcal{D}_3(k)$ belongs to the class $\sM \subset \mathcal{D}_3(k)$ of \emph{3-majority dynamics} if its function $f(x_1,x_2,x_3)$ has the clear-majority and the uniform properties.
\end{definition}

\begin{definition}[$(s,\varepsilon)$-plurality consensus solver]
We say that a protocol $\mathcal{P}$ is an \emph{$(s,\varepsilon)$-solver} (for the plurality consensus problem) if for every initial 
$s$-biased configuration $\mathbf c$, when running $\mathcal{P}$, with probability at least $1-\varepsilon$ there is a round $t$ by which all
nodes gets the plurality color of $c$.
\end{definition}

\noindent
Let us observe that, by definition of $h$-dynamics, any monochromatic configuration is an absorbing state of the relative Markovian process.
Moreover, the smaller $s$ and $\varepsilon$ the better an $(s,\varepsilon)$-solver is; in other words, if a dynamics is an $(s,\varepsilon)$-solver then it is also an $(s',\varepsilon')$-solver for every $s' \geqslant s$ and $\varepsilon' \geqslant \varepsilon$.
In Section~\ref{sec::UB}, we showed that any dynamics in $\sM$ is a $( \Theta( \sqrt{ \min\{ 2k  , (n/\log n)^{1/3}  \}  n \log n}), \Theta(1/n) )$-solver in $\mathcal{D}_3$. 
We can now state the main result of this section.

\begin{theorem}[properties of good solvers] \label{thm::LBgen} 
    Given a protocol $\mathcal{P}$, the following hold:
    \begin{enumerate}
        \item[(a)] If $\mathcal{P}$ is an $(n/4,1/4)$-solver in
            $\mathcal{D}_3$, then its $f$ must have the clear-majority property.
        \item[(b)] A constant $\eta > 0$ exists such that, if $\mathcal{P}$ is
            an $(\eta \cdot n,1/4)$-solver, then its $f$ must have the uniform
            property.
    \end{enumerate}
\end{theorem}

\noindent
The above theorem also provides the clear reason why some dynamics can solve consensus but cannot solve plurality consensus in the non-binary case.
A relevant example is the \emph{median} dynamics studied in~\cite{DGMSS11}: it has the clear-majority property but not the uniform one.

For readability sake, we split the proof of the above theorem in two technical lemmas: in the first one, we show the first claim about clear majority while in the second lemma we show the second claim about the uniform property.

\begin{lemma}[clear majority] \label{lm:clearmaj}
If a protocol $\mathcal{P} \in \mathcal{D}_3$ is an $(n/4,1/4)$-solver, then it chooses the majority color every time there is a triple with a clear majority.
\end{lemma}
\proof
For every triple of colors $(x_1,x_2,x_3) \in [k]^3$ that has a clear majority, let us define $\delta(x_1,x_2,x_3)$ to be $1$ if protocol $\mathcal{P}$ behaves like the majority protocol over triple $(x_1,x_2,x_3)$ and $0$ otherwise. Consider an initial configuration with only two colors, say red (r) and blue (b), with $c_r$ red nodes and $c_b = n-c_r$ blue nodes. Let us define $\Delta_r$ and $\Delta_b$ as follows
\begin{align*}
\Delta_r & = \delta(r,r,b) + \delta(r,b,r) + \delta(b,r,r), \\
\Delta_b & = \delta(b,b,r) + \delta(b,r,b) + \delta(r,b,b).
\end{align*}
We can write the probability that a node chooses color red as
\begin{align}\label{eq:probred}
p(r) & = \left(\frac{c_r}{n}\right)^3 + \left( \frac{c_r}{n} \right)^2
    \frac{c_b}{n} \cdot \Delta_r + \left(\frac{c_b}{n}\right)^2 \frac{c_r}{n}
    \left(3 - \Delta_b \right) \nonumber \\
& = \frac{c_r}{n^3}\left( c_r^2 + c_b \left( c_r \Delta_r - c_b \Delta_b
    \right) + 3 c_b^2 \right).
\end{align}
Observe that for a majority protocol we have that $\Delta_r = \Delta_b = 3$. In what follows we show that if this is not the case then there are configurations where the majority color does not increase in expectation. We distinguish two cases, case $\Delta_r \neq \Delta_b$ and case $\Delta_r = \Delta_b$.

\smallskip\noindent
\underline{Case $\Delta_r \neq \Delta_b$:} Suppose w.l.o.g. that $\Delta_r < \Delta_b$, and observe that since they have integer values it means $\Delta_r \leqslant \Delta_b - 1$. Now we show that, if we start from a configuration where the red color has the majority of nodes, the number of red nodes decreases in expectation. By using $\Delta_r \leqslant \Delta_b - 1$ in (\ref{eq:probred}) we get
\begin{equation}\label{eq:probred2}
p(r) \leqslant \frac{c_r}{n^3} \left( c_r^2 + c_b (c_r - c_b) \Delta_b - c_r c_b + 3 c_b^2 \right).
\end{equation}
If the majority of nodes is red then $c_r - c_b$ is positive, and since $\Delta_b$ can be at most $3$ from (\ref{eq:probred2}) we get
\begin{equation}\label{eq:probred3}
p(r) \leqslant \frac{c_r}{n^3}\left( c_r^2 + 2 c_r c_b \right).
\end{equation}
Finally, if we put $c_r = n/2 + s$ and $c_b = n/2 - s$, for some positive $s$, in (\ref{eq:probred3}), we get that
\begin{equation}\label{eq:probredcase1}
p(r) \leqslant \frac{c_r}{n^3} \left( \frac{3}{4}n^2 + (n-s)s \right) \leqslant \frac{c_r}{n}.
\end{equation}

\smallskip\noindent
\underline{Case $\Delta_r = \Delta_b$:} When $\Delta_r = \Delta_b$, observe that if the protocol is not a majority protocol then it must be $\Delta_r = \Delta_b \leqslant 2$. Hence, if we start again from a configuration where $c_r \geqslant c_b$, from (\ref{eq:probred}) we get that
\begin{equation}\label{eq:probredcase2}
p(r) \leqslant \frac{c_r}{n^3} \left( c_r^2 + 2 c_b (c_r - c_b) + 3 c_b^2 \right) = \frac{c_r}{n}.
\end{equation}

\smallskip\noindent
In both cases, for any protocol $\mathcal{P}$ that does not behave like a majority protocol on triples with a clear majority, if we name $X_t$ the random variable indicating the number of red nodes at round $t$, from (\ref{eq:probredcase1}) and (\ref{eq:probredcase2}) we get that $\Expec{}{X_{t+1} \,|\, X_t} \leqslant X_t$, hence $X_t$ is a supermartingale. Now let $\tau$ be the random variable indicating the first time the chain hits one of the two absorbing states, i.e.
\[
\tau = \inf\{ t \in \mathbb{N} \,:\, X_t \in \{0,n\} \}.
\]
Since $\Prob{}{\tau < \infty} = 1$ and all $X_t$'s have values bounded between
$0$ and $n$, from the martingale stopping theorem\footnote{See e.g. Chapter 17
in~\cite{lpwAMS08} for a summary of martingales and related results} we get
that $\Expec{}{X_\tau} \leqslant \Expec{}{X_0}$. If we start from a
configuration that is $n/4$-unbalanced in favor of the red color, we have
that $X_0 = n/2 + n/8$, and if we call $\varepsilon$ is the probability
that the process ends up with all blue nodes we have that $\Expec{}{X_\tau}
= (1-\varepsilon) n$. Hence it must be $(1-\varepsilon)n \leqslant n/2 +
n/8$ and the probability to end up with all blue nodes is $\varepsilon
\geqslant 5/8 > 1/4$. Thus the protocol is not a $(n/4,1/4)$-solver.
\qed

\begin{lemma}[uniform property]\label{lm::3distinct}
A constant $\eta > 0$ exists such that, if $\mathcal{P}$ is an $(\eta n,1/4)$-solver, then its $f$ must have the uniform property.
\end{lemma}
\proof
Thanks to the previous lemma, we can assume that $f$ has the clear-majority property but a triple $(r,g,b)$ exists such that $\delta_r < \max\{ \delta_g,\delta_b \}$.
Let us  start the process with the following initial configuration having only the above 3 colors and then show that the process w.h.p. will not converge to the plurality color $r$:
\[
\mathbf{c} = (c_r,c_g,c_b) = (n/3 + s,\, n/3,\, n/3 - s)
\quad \mbox{ where }
 s = \Theta(\sqrt{n\log n}).
\]
We consider the ``hardest'' case where $\delta_r = 1$: the case $\delta_r = 0$ is simpler since in this case, no matter how the other $\delta's$ are distributed, it is easy to see that the r.v. $c_r$ will decrease exponentially to $0$ starting from the above configuration.

\smallskip
\noindent
\textbf{- Case $\delta_r = 1$, $\delta_g = 3$, and $\delta_b = 2$} (and color-symmetric cases).
Starting from the above initial configuration, we can compute the probability $p(r) = \probc{X_v= r}{C = \mathbf c}$ that a node gets the color $r$.
\begin{align*}
p(r) & =  \left( \frac{c_r}{n}\right)^3 + 3 \left( \frac{c_r}{n} \right)^2
    \frac{n-c_r}{n} + \frac{c_r c_g c_b}{n^3} \\
& =  \frac{n+3s}{3 n^3} \left( \left(  \frac n3 + s\right)^2 + 3  \left(
    \frac n3 + s\right)\left(\frac 23 n - s \right) + \left( \frac n3\right)\left(
    \frac n3 - s \right) \right).
\end{align*}

\noindent
After some easy calculations, we get
\begin{equation*} \label{eq:prr}
p(r) = \frac 8 {27} \, \left(  1 + O\left(  \frac sn \right)\right).
\end{equation*}

\noindent
As for $p(g)$, by similar calculations, we obtain the following bound

\begin{equation*} \label{eq:prg}
p(g) = \frac {10} {27} \, \left(  1 - O\left(  \frac {s^2}{n^2} \right)\right).
\end{equation*}
From the above two equations, we   get  the following bounds on the expectation of the r.v.'s $X^{r}$ and $X^g$ counting the nodes colored with $r$ and $g$, respectively (at the next round).
\begin{align*}
    \expec{X^r}{\mathbf{C}= \mathbf c} & \leqslant \frac 8{27}\, n + O(s) \ \mbox{ and } \\
    \expec{X^g}{\mathbf{C} = \mathbf c} & \geqslant  \ \frac {10}{27} \, n - O\left(\frac{s^2}{n} \right).
\end{align*}
By a standard application of Chernoff's Bound, we can prove that, if $s \leqslant \eta n$ for a sufficiently small $\eta>0$, the initial value $c_r$ will w.h.p. decreases by a constant factor, going much below the new plurality $c_g$. Then, by applying iteratively the above reasoning we get that the process will not converge to $r$, w.h.p. 

\smallskip
\noindent
\textbf{- Case $\delta_r = 1$,  $\delta_g =  4$, and $\delta_b = 1$} (and color-symmetric cases).
In this case it is even simpler to show that w.h.p., starting from the same initial configuration considered in the previous case, the process will not converge to color $r$.
\qed

\subsection{A lower bound for h-plurality} \label{ssec::LB-h}
In Subsection~\ref{sec::LBM}, we have shown that the $3$-majority protocol
takes $\Theta(k \log n)$ rounds w.h.p. to converge in the worst case. A natural
question is whether by using the $h$-plurality protocol, with $h$ slightly
larger than $3$, it is possible to significantly speed-up the process. We prove
that this is not the case.

Let us  consider a set of $n$ nodes, each node colored with one out of $k$ colors. The $h$-plurality protocol works as follows:
\begin{quote}
\textit{At every round, every node picks $h$ nodes uniformly at random (including itself and with repetitions) and recolors itself according to the plurality of the colors it sees (breaking ties u.a.r.)}
\end{quote}
Let $j \in [k]$ be an arbitrary color, in the next lemma we prove that, if the number of $j$-colored nodes is smaller than $2 n/k$ and if $k/h = \mathcal{O}(n^{(1-\varepsilon) / 4})$, then the probability that the number of $j$-nodes increases by a factor $(1 + h^2 / k)$ is exponentially small.

\begin{lemma}\label{lemma:hincrate}
Let $\mathbf{c} = (c_1, \dots, c_k)$ be a configuration and let $j \in [k]$ be a color such that $(n/k) \leqslant c_j \leqslant 2(n/k)$. If $k/h = \mathcal{O}(n^{(1-\varepsilon) / 4})$ then it holds that
\[
\Prob{}{\left. C_j^{(t+1)} \geqslant \left( 1 + \frac{h^2}{k} \right) c_j \,\right|\, \mathbf{C}^{(t)} = \mathbf{c}} \leqslant e^{-\Theta(n^\varepsilon)}.
\]
\end{lemma}
\proof
Consider a specific node, say $u \in [n]$, let $N_j$ be the number of
$j$-colored nodes picked by $u$ during the sampling stage of the $t$-th round
and let $Y$ be the indicator random variable of the event that node $u$ chooses
color $j$ at round $t+1$. We give an upper bound on the probability of the
event $Y=1$ by conditioning it on $N_j = 1$ and $N_j \geqslant 2$ (observe that
if $N_j = 0$ node $u$ cannot choose $j$ as its color at the next round)
\begin{equation}\label{eq:ubprob1}
\Prob{}{Y_u = 1} \leqslant \Prob{}{Y_u = 1 \,|\, N_j = 1} \Prob{}{N_j = 1} + \Prob{}{N_j \geqslant 2}.
\end{equation}
Now observe that
\begin{itemize}
\item $\Prob{}{Y_u = 1 \,|\, N_j(u) = 1} \leqslant 1/h$ since it is exactly $1/h$ if all other sampled nodes have distinct colors and it is $0$ otherwise;
\item $\Prob{}{N_j = 1} \leqslant h {c_j} / {n}$ since it can be bounded by the probability that at least one of the $h$ samples gives color $j$;
\item $\Prob{}{N_j \geqslant 2} \leqslant \binom{h}{2} {c_j^2} / {n^2}$ since it is the probability that a pair of sampled nodes exist with the same color $j$. 
\end{itemize}
Hence, in~(\ref{eq:ubprob1}) we have that
\[
\Prob{}{Y = 1} \leqslant \frac{c_j}{n} + \frac{h^2}{2} \cdot \frac{c_j^2}{n^2}.
\]
Thus, for the expected number of $j$-colored nodes at the next round we get
\[
\Expec{}{C_j^{(t+1)} \,|\, \mathbf{C}^{(t)} = \mathbf{c}} \leqslant c_j + \frac{h^2}{2n} c_j^2 = c_j \left( 1 + \frac{h^2}{2n} c_j \right) \leqslant c_j \left( 1 + \frac{h^2}{k} \right),
\]
where in the last inequality we used the hypothesis $c_j \leqslant 2(n/k)$.
Since $C_j^{(t+1)}$ conditional on $\{ \mathbf{C}^{(t)} = \mathbf{c} \}$ is a sum of $n$ independent Bernoulli random variables, from the Chernoff bound (Lemma~\ref{lemma:cb} with $\lambda = c_j h^2 / k)$,  we finally get
\begin{align*}
\Prob{}{\left. C_j^{(t+1)} \geqslant c_j \left( 1 + 2 \frac{h^2}{k} \right) \,\right|\, \mathbf{C}^{(t)} = \mathbf{c} }
& \leqslant \exp\left(- \frac{2 (c_j h^2/k)^2}{n}\right) \\
& \leqslant \exp\left(- \Omega(n^\varepsilon)\right),
\end{align*}
where in the last inequality we used $c_j \geqslant n/k$ and $k/h = \mathcal{O}(n^{(1-\varepsilon) / 4})$.
\qed

\medskip \noindent
By adopting a similar argument to that used for proving Theorem~\ref{theorem:lowerbound}, we can get 
 a lower bound $\Omega(k/h^2)$ on the completion time of the $h$-plurality.

\begin{theorem}\label{theorem:h-lb}
Let $\mathbf{C}^{(t)}$ be the random variable indicating the configuration at round
$t$ according to the $h$-plurality protocol and let $\tau = \inf\{ t \in
\mathbb{N} \,:\, \mathbf{C}^{(t)} \mbox{ is monochromatic} \}$. If the initial
configuration $\mathbf{c} = (c_1, \dots, c_k)$ is such that $\max \{ c_j \,:\,
j = 1, \dots, k \} \leqslant {3n} / {(2k)}$ then $\tau =
\Omega(k / h^2)$ w.h.p.
\end{theorem}
\proof
Since in the initial configuration for any color $j \in [k]$ we have that $c_j \leqslant 3n /
(2k)$, from Lemma~\ref{lemma:hincrate} it follows that the number of
nodes supporting the plurality color increases at a rate smaller than $(1 +
2h^2 / k)$ with probability exponentially close to $1$. This easily implies a
recursive relation of the form $C_j^{(t+1)} \leqslant \left( 1 + 2h^2 / k \right)
C_j^{(t)}$ which, in turn, gives 
\[
C_j^{(t)} \leqslant \left( 1 + \frac{2h^2}{k} \right)^t C_j^{(0)} 
    \leqslant \left( 1 + \frac{2h^2}{k}
    \right)^t \frac{3n}{2k}.
\]
Thus, for $t < {k}/{h^2} \log ( 4/3 )$, w.h.p. we have that
\[ 
    C_j^{(t)} \leq \frac {3n}{2k} \left( 1 + \frac{2h^2 }{ k }\right)^t < \frac{2n}k ,
\]
concluding the proof.
\qed

\subsection{On the initial bias} \label{app:initbias}
In this section, we show that  there are initial configurations 
with  bias  $s = \mathcal{O}(\sqrt{kn})$ for which the bias  
decreases in a single round with constant probability. This shows
that under initial imbalances of this magnitude, it seems unlikely 
that one can prove bounds as those shown in Section~\ref{sec::UB}, at 
least with high probability.

\begin{lemma}\label{le:minimb}
Assume $k \ge 4$. For any value $s\le{\sqrt{kn}} / {6}$ of the 
initial bias, there are initial configurations $\mathbf{c}$ such 
that, for any fixed color $j \neq 1$ we have: 
\begin{align*}\
&\probb{C^{(1)}_1 - C^{(1)}_j < s|\mathbf{C}^{(0)} = \mathbf{c}} \ge\frac{1}{16e}.
\end{align*}
\end{lemma}
\proof
We consider an initial configuration $\mathbf{c}$ in which we have 
$k$ colors. Let $x = (n-s)/k$ (we neglect integer parts for the sake of the analysis).
We let $c^{(0)}_1 = x + s$ 
and $c^{(0)}_j = x$, for $j \ne 1$ and we further assume that $s\leqslant x$. 
Considered any fixed $j \neq 1$, we next prove that $C^{(1)}_1 - C^{(1)}_j < s$ 
with constant probability. 

\noindent
The outline of the proof is as follows: We first  show that 
$\Expec{}{C^{(1)}_1 \,| \, \mathbf{C}^{(0)} = \mathbf{c}} - \Expec{}{C^{(1)}_j \,|\,\mathbf{C}^{(0)} = \mathbf{c}}\leqslant s + {3xs} / {n}$, then we observe  
that with constant probablity $\Expec{}{C^{(1)}_1 \,|\, \mathbf{C}^{(0)} = \mathbf{c}}$ is not above its average. 
Finally,
we prove  that $C^{(1)}_j > \Expec{}{C^{(1)}_j \,|\, \mathbf{C}^{(0)} = \mathbf{c}} + {3xs} / {n}$ with constant 
probability, whenever $s \le {\sqrt{kn}} / {6}$, which concludes the 
proof of the lemma. 

\noindent
To begin with, from Lemma~\ref{lemma:3maj-expected}, we 
easily get the following derivations:
\begin{align*}
 \Expec{}{C^{(1)}_1 \,|\, \mathbf{C}^{(0)} = \mathbf{c}} & = x + s + \frac{x^2}{n} +
 \frac{2xs + s^2}{n} - \frac{x + s}{n^2} \gamma \quad  \mbox{and} \\
 \Expec{}{C^{(1)}_j \,|\, \mathbf{C}^{(0)} = \mathbf{c}} & = x + \frac{x^2}{n} - \frac{x}{n^2}\gamma,
\end{align*}
where $\gamma = \sum_h x^2_h = nx + xs + s^2$.

\noindent
Next, from the definition of  $\gamma$ we get
\begin{align*}
\Expec{}{C^{(1)}_1 \,|\, \mathbf{C}^{(0)} = \mathbf{c}} & = x + s + \frac{x^2}{n} +
\frac{2 x s + s^2}{n} - \frac{x + s}{n^2}(n x + x s + s^2) \\
    &= x + s + \frac{x s}{n} +
\frac{s^2}{n} - \frac{s}{n^2}(x + s)^2\\
& \leqslant x + s + \frac{2 x s}{n} - \frac{s}{n^2}(x + s)^2 < x + s + \frac{2 x s}{n},
\end{align*}
where the third inequality follows by assuming $s \leqslant x$. Analogously we have:
\begin{align}
\Expec{}{C^{(1)}_j \,|\, \mathbf{C}^{(0)} = \mathbf{c}} & = x + \frac{x^2}{n} -
    \frac{x}{n^2}(n x + x s + s^2) \nonumber\\
& = x - \frac{x s}{n}\cdot\frac{x+s}{n} \geqslant x - \frac{2 x^2 s}{n^2}\label{eq:cj},
\end{align}
where to derive the last inequality we again use $s \leqslant x$. As a 
consequence we can write:
\begin{equation*}
\Expec{}{C^{(1)}_1 \,|\, \mathbf{C}^{(0)} = \mathbf{c}} - \Expec{}{C^{(1)}_j \,|\, \mathbf{C}^{(0)}
    = \mathbf{c}} \le s + \frac{2 x s}{n} - \frac{2 x^2 s}{n}\leqslant s +
    \frac{3 x s}{n},
\end{equation*}
where the second inequality holds whenever $x \le n/2$, which is our 
case.

\noindent
For convenience sake, let us name for any $j \in [k]$
\[
\mu_j := \Expec{}{C^{(1)}_j \,|\, \mathbf{C}^{(0)} = \mathbf{c}}
\] 
We note that 
\[
\probb{C^{(1)}_1 - C^{(1)}_j < s \, | \, \mathbf{C}^{(0)} = \mathbf{c}} \ge \probb{C^{(1)}_1 <\mu_1\pand 
C^{(1)}_j \ge \mu_j + \frac{3 x s}{n} \, | \, \mathbf{C}^{(0)} = \mathbf{c}}.
\]

\begin{fact}\label{fa:negass}
The following holds:
\begin{multline*}
    \probb{ C^{(1)}_1 < \mu_1 \pand C^{(1)}_j \ge \mu_j + \frac{3 x s}{n} 
    \, \Big | \,  \mathbf{C}^{(0)} = \mathbf{c}}\\
    \ge\probb{ C^{(1)}_1 < \mu_1 \, \Big | \,  \mathbf{C}^{(0)} = \mathbf{c}}
    \probb{C^{(1)}_j \ge \mu_j + \frac{3 x s}{n}\, \Big | \,  \mathbf{C}^{(0)} = \mathbf{c}}.
\end{multline*}
\end{fact}

\begin{quote}
\begin{proof}
We have:
\begin{align*}
    &\probb{C^{(1)}_1 < \mu_1 \pand C^{(1)}_j \ge \mu_j + \frac{3 x s}{n}
        \, \Big | \,  \mathbf{C}^{(0)} = \mathbf{c}} \\
    &= \probb{C^{(1)}_1 < \mu_1 
        \, \Big | \, C^{(1)}_j \ge \mu_j + \frac{3 x s}{n}}
        \probb{C^{(1)}_j \ge \mu_j + \frac{3 x s}{n} 
        \, \Big | \, \mathbf{C}^{(0)} = \mathbf{c}} \\
    &= \left(1 - \probb{C^{(1)}_1 \geq \mu_1 
        \, \Big | \, C^{(1)}_j \ge \mu_j + \frac{3 x s}{n}}\right)
        \probb{C^{(1)}_j \ge \mu_j + \frac{3 x s}{n}
        \, \Big | \,  \mathbf{C}^{(0)} = \mathbf{c}} \\
    &= \probb{C^{(1)}_j \ge \mu_j + \frac{3 x s}{n}\, \Big | \, \mathbf{C}^{(0)} 
        = \mathbf{c}} - \probb{C^{(1)}_1 \geq \mu_1 \pand C^{(1)}_j \ge \mu_j + \frac{3 x s}{n}
        \, \Big | \,  \mathbf{C}^{(0)} = \mathbf{c}}\\
    &\geqslant \probb{C^{(1)}_j \ge \mu_j + \frac{3 x s}{n}
        \, \Big | \,  \mathbf{C}^{(0)} = \mathbf{c}} 
        - \probb{C^{(1)}_1 < \mu_1 \, \Big | \,  \mathbf{C}^{(0)} = \mathbf{c}}
        \probb{C^{(1)}_j \ge \mu_j + \frac{3 x s}{n}\, \Big | \,  \mathbf{C}^{(0)} = \mathbf{c}} \\
    &= \probb{C^{(1)}_j \ge \mu_j + \frac{3 x s}{n}\, \Big | \,  \mathbf{C}^{(0)} = \mathbf{c}} \\
    & \qquad - \left(1 - \probb{C^{(1)}_1 < \mu_1 \, \Big | \, \mathbf{C}^{(0)} = \mathbf{c}}\right)
        \probb{C^{(1)}_j \ge \mu_j + \frac{3 x s}{n}\, \Big | \, \mathbf{C}^{(0)} = \mathbf{c}}\\
    &= \probb{C^{(1)}_1 < \mu_1 \, \Big | \, \mathbf{C}^{(0)} = \mathbf{c}}
        \probb{C^{(1)}_j \ge \mu_j + \frac{3 x s}{n}\, \Big | \,  \mathbf{C}^{(0)} = \mathbf{c}},
\end{align*}
where the fourth inequality follows from \cite[Proposition 3, claim
(-OD)]{dubhashi1998balls}. In particular, $\left\{C^{(1)}_1 \geq \mu_1 \right\}$ 
and $\left\{ C^{(1)}_j \ge \mu_j + {3 x s} / {n}\right \}$ are the
events that the numbers of balls thrown independently at random into two
distinct bins both exceed some given thresholds.
\qed
\end{proof}
\end{quote}

\smallskip

\begin{fact}\label{fa:binomial}
The following holds:
\[
    \probb{C^{(1)}_1 < \mu_1 \,|\, \mathbf{C}^{(0)} = \mathbf{c}} > \frac{1}{4}.
\]
\end{fact}

\begin{quote}
\proof
Set $X = n - C^{(1)}_1$. Clearly, $X$ is distributed as a binomial $B(n, p)$, where $p = 1 - p_1$, with $p_1$ the probability that the generic node recolors itself with color $1$. Clearly, $p > {1}/{n}$ as long as the number of colors is not too large (in the order of $n$). Then we have:
\begin{align*}
&& \probb{C^{(1)}_1 < \mu_1 \,|\, \mathbf{C}^{(0)} = \mathbf{c}} = 
\probb{X \ge \Expec{}{X} \,|\, \mathbf{C}^{(0)} = \mathbf{c}}>\frac{1}{4},
\end{align*}
where the second inequality follows from \cite[Theorem 1]{greenberg2013tight}.
\qed
\end{quote}

\smallskip

\noindent
We finally apply Theorem \ref{thm:revchernoff} to $C^{(1)}_j$ and we have:
\begin{equation}\label{eq::expC}
 \probb{C^{(1)}_j > \mu_j + \frac{3 x s}{n} \,|\, \mathbf{C}^{(0)} = 
 \mathbf{c}} \ge \frac{1}{4} e^{-\frac{18 x^2 s^2}{n^2 \mu_j}} \ge \frac{1}{4} e^{-\frac{18 x s^2}{n^2 - 2xs}}
\, \geqslant \, \frac 1{4e},
\end{equation}
where the second inequality follows from \eqref{eq:cj}
and the third one holds since $s \leqslant \sqrt{kn}/6$ and recalling 
that $x \leq n/k$.
Finally, from  Fact \ref{fa:binomial} and \eqref{eq::expC}, we get   the claim.
\qed

\section{ Open Questions} \label{sec::conc}

A general open question on the plurality consensus problem is whether an
\emph{efficient} dynamics exists that achieves plurality consensus in
polylogarithmic time for any function $k=k(n)$. By \emph{efficient} dynamics
for our adopted model, we mean any dynamics that requires small (i.e.
$\mathcal{O}(\log n)$) memory, small random samples, and small message size.  
          
 A more specific question about  our simple distributed model is to explore the
 case in  which   the initial bias $s$ is smaller than the lower bound assumed
 in our analysis (i.e. $s  \geqslant c \sqrt{ \min\{ 2k  , (n/\log n)^{1/3}  \}
 \, n \log n}$). Notice that when $k$ is polylogarithmic, we  required a bias
 which is only a polylogarithmic factor larger than the standard deviation
 $\Omega(\sqrt n)$: the latter is a lower bound for the initial bias  to
 converge (w.h.p.) to the plurality color.  As for    larger $k$, we did not
 derive any stronger bound on the required bias,  however, in Subsection \ref{app:initbias}, we have shown  
 some  initial configurations with  bias  $s = \mathcal{O}(\sqrt{kn})$ for which
 the initial bias  \emph{decreases} in a single round with constant
 probability. This   result implies that, when the initial bias $s$  is
 ``slightly'' smaller than ``ours'', the process may be \emph{non-monotone}
 w.r.t. the bias function $s(t)$. The fact that $s(t)$ is an
 increasing function played  a key-role in the proof of  our upper bound. So,
 under such a weaker assumption, if any  upper bound similar to ours might  be
 proved then a much more complex argument (departing from ours) seems to be
 necessary. 
 
In this work, we were   interested in deriving sufficient conditions under
which the $h$-plurality dynamics converges in polylogarithmic time. A further
interesting open question is to derive conditions on the parameters $k$, $s$,
and $h$ under which this dynamics converges very fast, i.e., in sublogarithmic
time.

\bibliographystyle{plain}
\bibliography{majority}

\bigskip

\appendix
\section{Useful Bounds} \label{sec::app}

\begin{lemma}[Chernoff bounds]\label{lemma:cb}
Let $X = \sum_{i=1}^n X_i$ where $X_i$'s are independent Bernoulli random variables and let $\mu = \Expec{}{X}$. Then,
\begin{enumerate}
\item For any $0 < \delta \leqslant 4$, $\probb{X > (1 + \delta)\mu} < e^{-\frac{\delta^2\mu}{4}}$;
\item For any $\delta \geqslant 4$, $\probb{X > (1 + \delta)\mu} < e^{-\delta\mu}$;
\item For any $\lambda > 0$, $\Prob{}{X \geqslant \mu + \lambda} \leqslant e^{-2 \lambda^2 / n}$.
\end{enumerate}
\end{lemma}

\begin{lemma}[Jensen inequality]\label{lemma:jensen}
Let $\phi \,:\, \mathbb{R} \rightarrow \mathbb{R}$ be a convex function and $x_1, \dots x_k \in \mathbb{R}$ be $k$ real numbers, then
\[
\phi\left( \frac{1}{k} \sum_{i=1}^k x_i \right) \leqslant \frac{1}{k} \sum_{i=1}^k \phi(x_i).
\]
\end{lemma}

In Section~\ref{app:initbias}, we use the following  
``reverse''-Chernoff bound \cite[Theorem 2]{mousavitight} \footnote{A 
number of pretty similar ``folklore'' results can be found in 
specialized mathematical forums, for example \url{http://cstheory.stackexchange.com/questions/14471/reverse-chernoff-bound}.}
   
\begin{theorem}[Reverse Chernoff bound]\label{thm:revchernoff}
Let $X$ be the sum of $m$ independent Bernoulli variables with probability $p\le{1} / {4}$ and let $\mu = pm$. Then, for any $t > 0$:
\[
\probb{X - \mu > t}\ge\frac{1}{4}e^{-\frac{2t^2}{\mu}}.
\]
\end{theorem}


\end{document}